\documentclass{jps-cp}

\title{Static and Fluctuating Magnetic Moments in the Ferroelectric Metal LiOsO$_3$}

\author{Franziska~K.~K.~\textsc{Kirschner},$^{1}$ Franz~\textsc{Lang},$^{1}$ Francis~L.~\textsc{Pratt},$^{2}$ Tom~\textsc{Lancaster}$^{3}$, Youguo~\textsc{Shi},$^{4,5}$ Yanfeng~\textsc{Guo},$^{1,4}$ Andrew~T.~\textsc{Boothroyd},$^{1}$ and Stephen~J.~\textsc{Blundell}.$^{1}$}

\inst{$^{1}$Department of Physics, University of Oxford, Clarendon Laboratory, Parks Road, Oxford, OX1 3PU, United Kingdom \\
$^{2}$ISIS Pulsed Neutron and Muon Facility, Rutherford Appleton Laboratory, Chilton, Oxfordshire OX11 0QX, United Kingdom \\
$^{3}$Durham University, Department of Physics, South Road, Durham, DH1 3LE, United Kingdom \\
$^{4}$Superconducting Properties Unit, National Institute for Materials Science, 1-1 Namiki, Tsukuba, Ibaraki 305-0044, Japan \\
$^{5}$Institute of Physics, Chinese Academy of Sciences, Beijing 100190, China}

\email{franziska.kirschner@physics.ox.ac.uk}

\recdate{June 21, 2017}

\abst{LiOsO$_3$ is the first example of a new class of material called a ferroelectric metal. We performed zero-field and longitudinal-field $\mu$SR, along with a combination of electronic structure and dipole field calculations, to determine the magnetic ground state of LiOsO$_3$. We find that the sample contains both static Li nuclear moments and dynamic Os electronic moments. Below $\approx 0.7$~K, the fluctuations of the Os moments slow down, though remain dynamic down to 0.08~K. We expect this could result in a frozen-out, disordered ground state at even lower temperatures.}

\kword{magnetism, muon-spin relaxation, ferroelectric metal}

\begin{document}
\maketitle

\section{Introduction}

The theoretical model of a ferroelectric metal was first postulated by Anderson and Blount in 1965 \cite{Anderson1965}, despite it being previously thought that the conduction electrons in metals would inhibit the electrostatic forces required to facilitate a ferroelectric instability \cite{Cochran1960, Lines2001}. Several promising candidates for a ferroelectric metal have been synthesised \cite{Sergienko2004, Tachibana2010, Kolodiazhnyi2010}, but it was not until the observation of a ferroelectric-like structural transition in LiOsO$_3$ \cite{Shi2013} that the existence of this new class of materials was confirmed. Around 140 K, a centrosymmetric ($R\bar{3}c$) to noncentrosymmetric ($R3c$) phase transition occurs in LiOsO$_3$ (which is equivalent to the ferroelectric transition in LiNbO$_3$ and LiTaO$_3$ \cite{Abrahams1973,Boysen1994}), involving a shift in the mean position of the $A$-site Li \cite{Xiang2014}. There have been several competing explanations for the mechanism behind this transition, including that it may be fundamentally order-disorder-like \cite{Shi2013, Liu2015} or alternatively that it is caused by the hybridization of electronic orbitals \cite{Giovannetti2014, LoVecchio2016}.

Resistivity measurements performed on single crystal samples show metallic behaviour over a very wide temperature range \cite{Shi2013}. Electronic structure calculations predict either metallic \cite{Shi2013} or semiconducting behaviour \cite{He2015}. Although the $t_{2g}$ band is half-filled, metallic behaviour may emerge either as a consequence of hybridization (which produces a highly asymmetric density of states) or due to Li vacancies. Though the experimental sample in Ref.~\cite{Shi2013} had formula Li$_{0.98}$OsO$_3$, it has been argued that the Li-vacancy level may need to be closer to a Li$_{0.94}$OsO$_3$ composition in order to allow the ferroelectric metallic phase to exist \cite{He2015}. It has also been suggested that strong electron-electron correlations bring LiOsO$_3$ close to a Mott-Hubbard transition, but these are not strong enough to produce insulating behaviour \cite{Giovannetti2014, LoVecchio2016}. 

Experimental evidence for the magnetic nature of LiOsO$_3$ remains limited. The susceptibility shows Curie-Weiss-like behaviour below the structural phase transition suggesting the presence of localised paramagnetic moments, but there is no indication of a magnetic ordering transition. Neutron powder diffraction measurements place an upper limit of $\approx 0.2~\mu_{\rm B}$ on the size of any ordered magnetic moments \cite{Shi2013}. Theoretical studies have ruled out Slater antiferromagnetism due to the nesting of Fermi levels \cite{Giovannetti2014}. The true magnetic ground state of LiOsO$_3$ has not yet been ascertained by experiment, with Density Functional Theory (DFT) calculations in Ref.~\cite{Giovannetti2014} indicating the presence of magnetic correlations and an instability towards G-type magnetic order. Reference~\cite{He2015} provides an alternative scenario, in which LiOsO$_3$ has a G-type antiferromagnetic (AFM) ground state. This prediction can be reconciled with the experimental observations through the introduction of Li defects, which reduce the Os magnetic moment and may be responsible for the destruction of magnetic order. In high-defect structures, it is possible that magnetism may be destroyed entirely. In the case that the ferroelectric transition were to be caused by band hybridization, however, the AFM state is disfavoured as it would result in an insulating ground state instead of a metallic one \cite{LoVecchio2016}.

In this paper, we present the results of muon-spin relaxation ($\mu$SR) experiments on a powder sample of LiOsO$_3$, in order to determine its magnetic properties. We then calculate the muon site using DFT+$\mu$ \cite{Moeller2013} and model the magnetic field experienced by the muon through dipole field calculations. Our results show the coexistence of disordered static Li nuclear moments with dynamic Os electronic moments. We also find that the Os electronic moments begin to slow down markedly at around 0.7~K, with some dynamics persisting down to 0.08~K. It is reasonable to expect that the dynamics may eventually freeze out completely at even lower temperatures, resulting in a ground state consisting of static, disordered Li nuclear moments and static, disordered Os electronic moments.

\section{Experimental details}

The polycrystalline sample of LiOsO$_3$ used in the experiments reported in this paper was the same as that studied previously by neutron diffraction \cite{Shi2013}. Details of the high pressure synthesis and physical characterisation of the sample are described in Ref.~\cite{Shi2013}.

$\mu$SR experiments \cite{Blundell1999, Yaouanc2011} were performed using a dilution refrigerator mounted on the MuSR spectrometer at the ISIS pulsed muon facility (Rutherford Appleton Laboratory, UK) \cite{King2013}. Zero-field (ZF) and longitudinal-field (LF) measurements were carried out in order to test for magnetic phases in the sample. Preliminary experiments were carried out using the low temperature facility (LTF) and general purpose spectrometer (GPS) at the Swiss Muon Source. All of the data were analyzed using WiMDA \cite{Pratt2000}.

\section{$\mu$SR experiments}

\begin{figure}[t]
	\includegraphics[width=\textwidth]{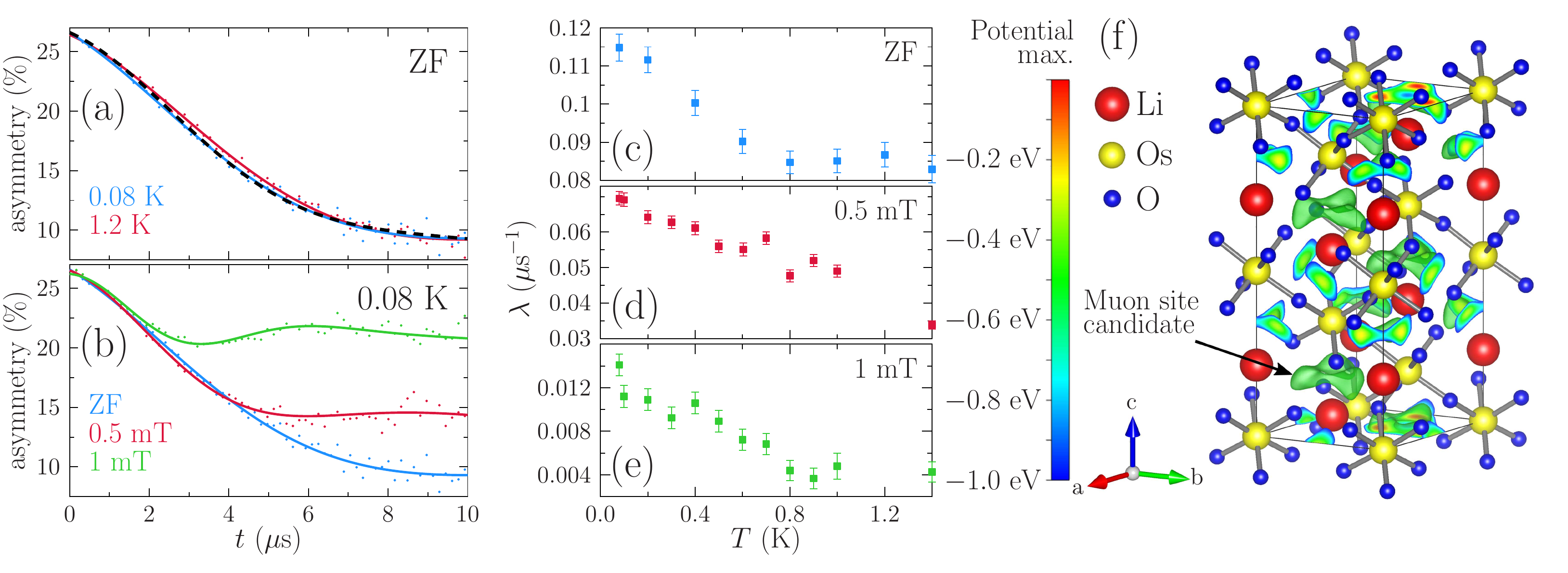}
	\caption{(a) Sample ZF-$\mu$SR spectra above and below the transition in LiOsO$_3$. Fits as in Eq.~\ref{KTLor} are also plotted. The simulated low-$T$ muon asymmetry, calculated as in Sec.~\ref{theory} is given by the dashed line. (b) ZF- and LF-$\mu$SR spectra at 0.08~K showing the suppression of relaxation. The temperature dependences of the relaxation parameter $\lambda$ in ZF, 0.5~mT, and 1~mT are shown in (c), (d), and (e) respectively. (f) Electrostatic Coulomb potential of LiOsO$_3$computed with DFT. The potential is shown on the surface of the unit cell up to $1.0$~eV below its maximum value, and a green isosurface is plotted within the unit cell at $0.5$~eV below the maximum.}
	\label{ZFLFfig}
\end{figure}

ZF measurements were carried out between temperatures $T$ of 0.08~K and 1.4~K, and example spectra are shown in Fig.~\ref{ZFLFfig}(a). No oscillations were seen in the forward-backward asymmetry spectra, and there were also no discontinuous jumps in either the initial or the baseline asymmetry, which suggests there is no long-range order inside the sample. There are also no large discontinuities in the observed relaxation, ruling out a transition to long-range order with a field distribution out of the range resolvable by the experiment. The data were fitted to a function $A(t)$ containing a zero-field Kubo-Toyabe function \cite{Kubo1967} with an additional Lorentzian relaxation, given by

\begin{equation} \label{KTLor}
A = A_0 \left( \frac{1}{3} + \frac{2}{3} \left( 1 - \Delta^2 t^2 \right) e^{ - \frac{1}{2} \Delta^2 t^2 } \right) e^{- \lambda t} + A_{\rm base},
\end{equation}
where $A_0$ and $A_{\rm base}$ are the initial and baseline amplitudes. This function is well-suited to describe systems with a mixture of static and dynamic spins \cite{Uemura1985}, where the static spin distribution is characterised by a width $\Delta/\gamma_{\mu}$ ($\gamma_{\mu} = 2 \pi \times 135.5~\rm{MHz T}^{-1}$ is the gyromagnetic ratio of the muon) and the dynamic distribution by a relaxation $\lambda=2\Delta_{\rm dyn}^2 / \nu$ where $\Delta_{\rm dyn}/\gamma_{\mu}$ is the width of the field distribution giving rise to the dynamic relaxation and $\nu$ is the spin fluctuation frequency.

It was found through fitting these parameters that the static portion of the relaxation remains constant with a frequency width given by $\Delta = 0.167(4)~\mu\rm{s}^{-1}$, corresponding to a field width of $\approx 0.2$~mT. The dynamic relaxation, on the other hand, increased significantly below $T \approx 0.7$~K, as shown in Fig.~\ref{ZFLFfig}(c), either due to a widening of the dynamic field distribution or the slowing down of fluctuations. In a preliminary experiment, we determined that the form of the relaxation is unchanged between 1.4~K and 120~K.

As the static relaxation is Kubo-Toyabe-like with a small $\Delta$, it is likely that these fields arise from nuclear moments \cite{Hayano1979}, namely from Li with $\mu \approx 3\mu_{\rm N}$ where $\mu_{\rm N}$ is the nuclear magneton. Previous $\mu$SR experiments on osmate compounds have observed fast fluctuating osmium spins which freeze out at low temperature \cite{Yan2014}. Therefore the dynamic component likely arises from Os$^{5+}$ electronic moments (5d$^3$, $S=3/2$).

To further probe the nature of the ZF relaxation, LF measurements were made in which an external magnetic field is applied along the direction of the initial muon spin polarization. LF-$\mu$SR is often used to distinguish between the static and dynamic fields experienced by the muon, as a longitudinally applied field of order $\Delta/\gamma_{\mu}$ rapidly quenches any static relaxation \cite{Yaouanc2011}. Fields of 0.5~mT and 1~mT were applied between temperatures of 0.08~K and 1.4~K, and sample spectra are shown in Fig.~\ref{ZFLFfig}(b). The data were fitted to a LF Kubo-Toyabe function \cite{Hayano1979} with an additional Lorentzian relaxation.

It is clear in Figs.~\ref{ZFLFfig}(b), (d), and (e) that the muon's polarization becomes decoupled in a longitudinal field, indicating the presence of static moments in the sample. From the value of the decoupling field it can be deduced that the static fields experienced by the muon are on the order of 0.1~mT (as the longitudinal field $B_{\rm LF} > 10 B_{\rm internal}$ for decoupling to occur \cite{Hayano1979}). It should be noted, however, that the onset of the freezing in $\lambda$ persists even at 1~mT (see Fig.~\ref{ZFLFfig}(e)), which suggests this transition is caused by unquenched dynamic moments.

\section{DFT+$\mu$ and dipole field calculations} \label{theory}

We have carried out simulations to explore the effect of magnetism from the Li and Os at the muon site. In order to establish the potential muon stopping sites, we employ DFT calculations to map out the electrostatic Coulomb potential of LiOsO$_3$ throughout its unit cell. The maxima of such a potential map have been a reliable approximation to the muon sites in previous more in-depth DFT+$\mu$ calculations \cite{Moeller2013, Foronda2015, MoellerThesis}. Additionally, we carried out relaxation calculations which allow for local distortions of the lattice caused by the presence of the muon.

The DFT calculations were conducted using the plane-wave program {\it Quantum Espresso}~\cite{Gianozzi2009}, and utilized the generalized gradient approximation (GGA) for the exchange-correlation functional~\cite{Perdew1996}. The ions were modelled with ultrasoft pseudopotentials~\cite{Rappe1990}, while a norm-conserving hydrogen pseudopotential was used to model the muon. The energy cutoffs for the wavefunction and the charge density were set to $80$\,Ry and $800$\,Ry, respectively, and the integration over the Brillouin zone was carried out using a $3\!\times\! 3\! \times \!3$ Monkhorst-Pack $k$-space grid~\cite{Monkhorst1976}. These parameters were found to yield well-converged results that predicted atomic positions and lattice parameters of the bulk material within 1.5\% of the experimentally observed values~\cite{Shi2013}. The electrostatic Coulomb potential was computed from the converged electron density and the resulting three-dimensional visualisations were created with the Vesta software~\cite{Momma2008}.

The Coulomb potential of LiOsO$_3$ obtained through the DFT calculations is plotted in Figure~\ref{ZFLFfig}(f) with respect to the global maximum of the potential. Large values of the Coulomb potential correspond to low energies needed to add a positive charge (such as the muon). Therefore, the local maxima of the potential can be used to identify likely stopping sites for a muon. Relaxation calculations, in which all the ions were free to move until the forces on them had converged to less than 10$^{-3}$~Ry/a.u., resulted in a single muon site at $(0.195, 0.338, 0.319)$ and a $1~\AA$ O--H like bond with the nearest O. This is in line with the approximate position we identified from the electrostatic potential. Furthermore, the relaxation calculations show that the muon induced distortions are negligibly small ($<0.09~\AA$) on the Os positions and only significantly affect the nearest Li ion, which is displaced by approximately $1~\AA$.

Using this muon site $r_{\mu}$, it is possible to calculate the expected field width $\Delta$ as given in the Kubo-Toyabe function in Eq.~\ref{KTLor}. For a system of randomly oriented nuclear spins with nuclear quantum number $I$ at positions $r_i$,
\begin{equation} \label{deltaEq}
\Delta^2 = \frac{2}{3}\gamma_{\mu}^2 \gamma_{\rm nucl}^2 \left( \frac{\mu_0}{4 \pi} \right)^2 I(I+1) \hbar^2 \sum_i{\frac{1}{\left( r_i - r_{\mu} \right)^6}},
\end{equation}
where $\gamma_{\rm nucl}$ is the gyromagnetic ratio of the nuclei \cite{Yaouanc2011}. By considering both $^6$Li and $^7$Li nuclear spins and taking a weighted average of their individual $\Delta$ values, we find $\Delta = 0.172~\mu \rm{s}^{-1}$, which is consistent with the experimental value. As Eq.~\ref{deltaEq} does not consider quadrupole moments, we conclude that there is a very weak muon-induced electric field gradient at the nearest-neighbour Li sites. It should be noted that $\Delta$ is very sensitive to the position of the muon and the nearest Li ions, and the quoted result is for the positions obtained from the DFT+$\mu$ calculations of structural relaxation around the muon site.

 By considering a system containing static Li nuclear moments and dynamic Os electronic moments, we calculated the resulting static and dynamic dipolar field distributions at the muon site. For high fluctuation rates $\nu$, the resulting muon polarization spectrum is modelled by a product of the relaxations from the static and dynamic field distributions, as described by Eq.~\ref{KTLor}. This relaxation function was calculated by simulating the fields at the muon site resulting from randomly oriented static Li nuclear moments, along with randomly oriented Os electronic moments of size $\approx 1.5 \mu_{\rm B}$ \cite{He2015} which were then subject to fluctuations. It was found that the experimental data is well-modelled by such a system, with the simulated low-$T$ ZF asymmetry plotted in Fig.~\ref{ZFLFfig}(a) almost coincident with the fit to Eq.~\ref{KTLor}.  From these simulations, it can be concluded that the field at the muon site is likely from static, disordered Li nuclear moments, and dynamic Os electronic moments with a field width $\Delta_{\rm dyn}/\gamma_{\mu}\approx~0.1$~T and fluctuation rate $\nu\sim~10^5$~MHz. We also find that an increase in this fluctuation rate models the high-$T$ data well, supporting the notion that the increase in ZF and LF relaxation rates below 0.7~K is caused by the slowing down of fluctuations of the Os electronic moments. From these simulations, we confirm that the relaxation observed in the experimental data is not due to long-range order in the sample.

\section{Conclusions}

We have carried out ZF- and LF-$\mu$SR measurements on a powder sample of LiOsO$_3$ and obtained new information on the magnetic character of the ground state. The ZF data indicates the presence of both static and dynamic magnetic moments, with the dynamic fluctuations slowing below 0.7~K. The LF measurements support this notion, with the transition persisting in a field of 1~mT. Using DFT+$\mu$, we calculated the muon site and computed the dipole fields at this point. We find that it is likely the muon experiences a static field originating from randomly oriented Li nuclei, as well as fluctuating fields from Os electronic moments, where the fluctuation frequency decreases below 0.7~K but remains non-zero at 0.08~K. At even lower temperatures, these fluctuations are likely to decrease further, which would result in a ground state likely consisting of static disordered nuclear Li moments, as well as frozen-out, disordered Os electronic moments.

\section{Acknowledgements}
We thank C.~Baines for experimental assistance. F.K.K.K. thanks Lincoln College Oxford for a doctoral studentship. Part of this work was performed at the Science and Technology Facilities Council (STFC) ISIS Facility, Rutherford Appleton Laboratory, and part at S$\mu$S, the Swiss Muon Source (PSI, Switzerland). This work is supported by EPSRC (UK) grant no.~EP/N023803/1.

\end{document}